\documentclass{PoS}
\usepackage{paralist}

\title{Data processing and online reconstruction}

\ShortTitle{Data processing and online reconstruction}

\author{\speaker{David Rohr} on behalf of the ALICE, ATLAS, CMS, and LHCb collaborations\\
        CERN\\
        E-mail: \email{drohr@cern.ch}}

\abstract{
In the upcoming upgrades for Run 3 and 4, the LHC will significantly increase Pb--Pb and pp interaction rates.
This goes along with upgrades of all experiments, ALICE, ATLAS, CMS, and LHCb, related to both the detectors and the computing.
The online processing farms must employ faster, more efficient reconstruction algorithms to cope with the increased data rates, and data compression factors must increase to fit the data in the affordable capacity for permanent storage.
Due to different operating conditions and aims, the experiments follow different approaches, but there are several common trends like more extensive online computing and the adoption of hardware accelerators.
This paper gives an overview and compares the data processing approaches and the online computing farms of the LHC experiments today in Run 2 and for the upcoming LHC Run 3 and 4.
}

\FullConference{Sixth Annual Conference on Large Hadron Collider Physics (LHCP2018)\\
                4--9 June 2018\\
                Bologna, Italy}

\begin{document}

\section{Introduction}

During the next two long shutdowns, the LHC will install significant upgrades.
After LS2 (Long Shutdown 2), the Pb--Pb interaction rate will increase from currently around~10\,kHz to~50\,kHz.
This will go along with a major upgrade of the ALICE~\cite{bib:alice} detector and of the ALICE computing infrastructure~\cite{bib:aliceupg}.
In parallel, the LHCb will also upgrade both their detector and their computing farm to utilize a larger fraction of the available luminosity~\cite{bib:lhcbupg}.
For ATLAS~\cite{bib:atlasupg} and CMS~\cite{bib:cmsupg}, the pp interaction rate will only increase that much.
After LS3 in contrast, the pp rate will increase five to seven times the current nominal interaction rate, accompanied by large upgrades of ATLAS and CMS.
Studies of future hadron collider experiments plan with even higher collision rates.

These upgrades pose enormous challenges for online data processing and reconstruction.
The required compute capacity will usually grow more than linearly, since the upgrades bring about new detector features like high granularity calorimeters for Run 4, and with them more compute intense reconstruction algorithms.
In addition, the tracking and ambiguity solving complexity increases at the high planed pile-up of~$\mu \approx 200$ for Run 4.
Finally, trigger algorithms become more complex and need more input from the online reconstruction.
As a consequence, all experiments invest heavily in their data processing, in both hardware and software, and investigate new ways like hardware accelerators.

\section{ALICE}

The ALICE~\cite{bib:alice} data-taking is currently limited by the drift detectors like the TPC (Time Projection Chamber), which require the hardware triggers to reduce the readout rate to O(1)\,kHz.
In parallel, events are large due to the abundance of hits in the TPC and in particular due to the out-of-bunch pile-up of around 20 collisions at~200\,kHz proton--proton collisions.
A higher interaction rate yields larger events and, due to limited read-out bandwidth, leads to fewer events stored in the end.
In the online processing during Run~2, the ALICE HLT (High Level Trigger) does not operate like a trigger as in the other experiments and as initially designed.
Instead, it employs sophisticated online compression of the TPC raw data, the main contributor to data size.
Thanks to the compression, ALICE can store all events read out and does not need to employ software triggers.
This online compression will become even more important for Run~3.

In Run~3, ALICE will perform a major Online Offline computing Upgrade in the scope of the O$^2$ project~\cite{bib:o2tdr}.
This comes with a change of paradigm.
Instead of independent online and offline processing, one fast but limited, and the other slow but precise, ALICE will run a common software in a large processing farm.
The processing will run in two phases.
The synchronous processing performs data compression, calibration, and some necessary reconstruction steps online during data-taking.
Compressed raw data is stored onto a disk buffer at a rate of up to 100\,GB/s, compared to more than~3\,TB/s of data coming from the detectors.
A post-processing step may run afterwards to create the final calibration.
When the LHC does not provide stable beam (during refill, machine development, technical stop, year end technical stop, ...), the asynchronous phase will reprocess the data from the disk buffer using the final calibration and create the final reconstruction output.
There is no separate online and offline software, but both phases run the same code with different settings and calibration on the same computing farm.

In parallel, ALICE will upgrade its detectors and switch from the triggered read-out at around 1\,kHz to continuous read-out of 50\,kHz minimum bias Pb--Pb events.
Consequently, the online farm must be able to process and store 50 times as many events as today online, with offline quality.
This will require improved data compression~\cite{bib:lhcp2017}, based on full TPC online reconstruction.
The TPC calibration~\cite{bib:lhcp2017} does not need the full reconstructed data set, such that tracking for the other detectors can be limited to around~10\% of the events during the synchronous phase.
Hence, the online farm must be able to cope with~50\,kHz Pb--Pb data-taking, and the TPC online tracking at 50\,kHz Pb--Pb defines the peak processing capacity the online farm must provide.
ALICE counts on GPU-accelerated processing~\cite{bib:chep2016gpu} to provide this capacity.
The TPC tracking will require in the order of 1000 modern GPUs.

\looseness=-1
Looking at Run 2 online and offline processing, the TPC reconstruction needs the most compute cycles, after the GPU speed-up for the TPC tracking, other steps become bottlenecks.
In the offline processing, around~10\% of the time is spent for TPC energy loss d$E$/d$x$ calculation.
However, the current algorithm is inefficient and this will be sped up using GPUs.
The TPC common mode and ion tail correction need~17\% of current CPU time, but will no longer be needed after the Run 3 detector upgrade.
The ITS tracking (Inner Tracking System consisting of 7 layers of silicon pixels in Run 3) consumes~33\% of the CPU resources due to the combinatorics in the high hit-density situation, and ALICE is developing a new GPU-accelerated Cellular Automaton-based ITS tracking.
The remaining reconstruction tasks currently need~17\% of the compute time and must be sped up as well in order to achieve the necessary overall speed-up of at least an order of magnitude.

\section{LHCb}

LHCb employs an online processing scheme similar to what ALICE envisions for Run 3 already in Run 2.
Their L0 hardware trigger reduces the event rate from the 30\,MHz bunch crossing rate down to~1\,MHz, which is fed into the HLT farm.
The HLT runs in two phases: HLT1 and HLT2.
The HLT1 runs tracking and inclusive triggers without PID, and stores accepted raw data and calibration data to a disk buffer.
The HLT2 runs the second phase with full offline quality reconstruction and the final trigger decision asynchronously.

One special feature is the TURBO stream, which stores reconstructed HLT data directly for analysis.
This enables preliminary results already 1 week after data-taking.

Run 3 will bring two changes.
First, the TURBO stream will become the default.
There will not be posterior offline reconstruction, except for a small fraction reconstructed for QA (Quality Assurance) reasons.
Second, LHCb will abandon the hardware trigger completely, and do full software processing at the bunch crossing rate.
They aim to collect~10 times more events per period, which will come from a 5 times higher luminosity and the 2 times more efficient software trigger.
A detailed comparison of the LHCb and the ALICE approach is given in Section~\ref{sec:comparison}.

\section{ATLAS}

The ATLAS trigger is based mainly on calorimeter and muon items.
The data is buffered in the readout (Run 2) or FELIX (Run 3) and the trigger operates on regions of interests (ROI).
In this way, only the relevant data is transferred saving bandwidth, and the full event is only read out in case of a final trigger accept.
The ATLAS high level trigger reads the data as needed, and typically requests~50 to~80\% of the L1 rate of around~95\,kHz.
An additional input to the HLT is the Fast Tracker (FTK), which is a hardware tracking system based on FPGAs and an associative memory.
A partial FTK system is under commissioning in Run 2 and the full system is aimed to be deployed for Run 3 and will provide tracking at the full L1 rate.

The HLT itself is based on commercial multi-core servers.
Migration of code to GPUs has not proven to be cost-effective yet.
For Run 3, ATLAS aims at a multi-threaded software for offline and for the HLT called AthenaMT.
The HLT outputs full events at a rate of 1 to 2\,kHz, which dominate the event rate to storage of up to 3\,GB/s.
In addition, partial events consisting mostly of HLT jet objects can be stored at higher rates up to 15 kHz and are used for instance for luminosity measurements and dark matter searches.

\looseness=-1
For Run 4, ATLAS will evolve on this concept, and raise the hardware trigger rate to~1\,MHz, with the option to upgrade to~4\,MHz if needed.
An additional time-multiplexed global trigger will have access to the full calorimeter granularity but will increase the trigger latency.
Full events will be buffered by a storage handler and the online farm will again use commercial servers.
GPU offloading is under evaluation.
The hardware tracker FTK will evolve into the baseline tracking (HTT) again based on associative memory.
When the LHC does not provide collision data, the HTT will be used for the Monte Carlo simulations.
It is foreseen to store full events at a rate of~10\,kHz.

\section{CMS}

CMS features similar rates compared to ATLAS, currently storing 1\,kHz of prompt triggers and 500\,Hz of parked trigger raw data from the HLT, and aiming for~7.5\,kHz for Run~4.
An overview is given in Tables~\ref{tab:comparison1} and~\ref{tab:comparison2}.
As for all experiments, the data size and compute capacity estimates increase significantly with the update.
With the current extrapolation for Run 4, assuming a flat budget and usual technological improvements, CMS is a factor 4 short, while there are still around 10 years to close the gap.

Currently, most of the compute time goes into tracking, in particular with higher pile-up.
Therefore, the most important computing optimization has been related to tracking.
CMS is working with Cellular Automaton tracking, which is a common trend in high energy physics, also followed by ALICE for instance.
This situation will change with Run 4 when new detectors like high granularity calorimeters will appear.
Other reconstruction steps will require significant compute resources as well, which multiplies the required optimization effort.

Therefore, CMS is looking into multiple directions including artificial intelligence with ma\-chine and deep learning.
Support for the respective APIs is being integrated into CMSSW.
It is planed to use GPUs for the high level trigger, while GPU usage is not so clear offline due to the rising heterogeneity.
Also the simulation needs to be sped up, where GeantV will play a role.
In order to fit both raw and reconstructed data into the available storage, efficient and compressed storage formats are investigated, where CMS is pioneering NanoAODs.

\section{Comparison}
\label{sec:comparison}

\begin{table}
\begin{tabular}{l|ll|ll}
\hline
& ALICE & (Pb--Pb) & LHCb & \\
& Run 2 & Run 3/4 & Run 2 & Run 3/4 \\
\hline
Luminosity & 10\,kHz & 50\,kHz & $4 \cdot 10^{32}$ & $2 \cdot 10^{33}$ \\
\hline
Hardware & 500\,Hz -- 2\,kHz & 50\,kHz & 1\,MHz & 30\,MHz \\
Trigger & & (continuous) & & (bunch crossing) \\
\hline
HLT Accepts & No rejection & No HLT & 12.5\,kHz & $>100$\,kHz \\
\hline
Raw data rate & 45\,GB/s & 3\,TB/s & 55\,GB/s & 4\,TB/s \\
into HLT & (zero-suppressed) & (no ZS) & & (zero-suppressed) \\
\hline
Data stored & $\sim 10$\,GB/s & up to 100\,GB/s & 0.6\,GB/s & 2--10 GB/s \\
\hline
Data buffer & $\sim 1$\,PB DAQ & $\sim 60$\,PB & $\sim 12$\,PB & $\sim 100$\,PB \\
& buffer to Tier0 & one year of & & two weeks of HLT1 \\
& & compressed data, & & accepted raw data, \\
& & up to 100\,GB/s & & 150 + 150\,GB/s \\
& & & & read + write \\
\hline
\end{tabular}
\caption{Comparison of the Online Compute Farms (ALICE and LHCb).}
\label{tab:comparison1}
\end{table}

\begin{table}
\begin{tabular}{l|ll|ll}
\hline
& ATLAS & & CMS & \\
& Run 2/3 & Run 4 & Run 2/3 & Run 4 \\
\hline
Luminosity & $2 \cdot 10^{34}$ & $7.5 \cdot 10^{34}$ & $2 \cdot 10^{34}$ & $7.5 \cdot 10^{34}$ \\
\hline
Hardware & 95\,kHz & 1\,MHz & 100\,kHz & 500--750\,kHz \\
Trigger & & can evolve to 4 & & \\
\hline
HLT Accepts & 1--2\,kHz & 10\,kHz & 1.5\,kHz & 5--7.5\,kHz \\
\hline
Raw data rate & 29\,GB/s HLT req. & 2.6\,TB/s HLT req. & 200\,GB/s & 3--5.5\,TB/s \\
into HLT & 260\,GB/s L1 & 5.2\,TB/s L1 & (event network) & (event network) \\
\hline
Data stored & 2.4\,GB/s & 50\,GB/s & 5\,GB/s & 32--61\,GB/s \\
\hline
Data buffer & 1.5\,TB & 36\,PB & 12\,TB & 171--333\,TB \\
& event buffer + & 48 hours + & ram disk, & events before \\
& 48 hours to Tier0 & L1 to HLT & events before & HLT, 60s \\
& & & HLT, 60s & \\
\end{tabular}
\caption{Comparison of the Online Compute Farms (ATLAS and CMS).}
\label{tab:comparison2}
\end{table}

Tables~\ref{tab:comparison1} and~\ref{tab:comparison2} compare the online compute farms of all 4 experiments.
After the upgrade, ALICE will take 50 times more events than before, but only minimum bias events.
The other experiments will all record around 10 times more events.
The ALICE online farm could optionally be used as a trigger during some time of the year, with pp data-taking at higher rate in the order of~1\,MHz.
While ALICE and LHCb go for full online processing in software, ATLAS and CMS will stick to hardware triggers as today.
However, ATLAS and CMS run at a much higher collision rate such that full online processing would be infeasible due to budget and power consumption.
ALICE and LHCb will employ a disk buffer to store data after the first-phase real time processing.
Their online computing farms will reprocess the data from the buffer a second time asynchronously when the LHC does not provide stable beam.
Interestingly, ALICE has the highest data rate to permanent storage.
However, this is limited to only a short time period, at the beginning of 50\,kHz Pb--Pb fills, only for few weeks in the year.
In the following, we compare the ALICE and LHCb approaches in more detail.
\begin{compactitem}
\item \textbf{General approach}: Both experiments will run without a hardware trigger with a two-stage processing and an intermediate disk-buffer, providing offline-quality results in the online compute farm.
\item \textbf{Input data rates}: The rates will be between 3 and 4\,TB/s, but the LHCb data is zero-suppressed, hence the LHCb raw rate is higher.
\item \textbf{Calibration}: Both experiments perform the calibration in the first phase such that the final calibration is available for the second phase processing.
  Partial calibration like LHCb velo alignment can be done online in phase one with a feedback loop.
\item \textbf{Event building}: Both experiments have a dedicated subset of nodes (FLPs for ALICE, DAQ for LHCb) that receive the detector data via optical links with the PCIe40 card.
  ALICE collects time frames of around 20\,ms and LHCb coalesces multiple events, in order to reduce the scheduling rate.\newline
    \textit{(Up until here, the approach is identical, the following points show the differences.)}
    {\setlength\itemindent{25pt} \item \textbf{ALICE} sends the data from the FLPs directly to the Event Processing Nodes (EPN), where events are merged, built, and reconstructed, distributing time frames in a round-robin-like fashion.}
    {\setlength\itemindent{25pt} \item \textbf{LHCb} plans to build the events inside the DAQ with a fast network, and then distribute them with a slow but broad network to the event filter nodes for reconstruction}
\item \textbf{Cluster}: ALICE plans with around 200 FLPs and 1500 GPU-equipped EPNs, LHCb plans with around 500 DAQ and 4000 event filter nodes.
\item \textbf{Software}: ALICE is building a completely new software stack based on message queuing that will also be used for simulation and analysis, while LHCb is evolving the software inside the Gaudi framework.
\item \textbf{Disk buffer}: Although the disk buffer will be of similar size featuring a similar data rate, the use case is very different.
    {\setlength\itemindent{25pt} \item \textbf{LHCb} will buffer up to 3 weeks of raw data accepted by the HLT1, around 3 days is the minimum needed for calibration.
      This allows them to use the LHC turnaround, technical stops, and machine development for asynchronous processing while the farm can run Monte Carlo simulations during the year end technical stop.
      Read and write speed are symmetric with around 150\,GB/s each.}
    {\setlength\itemindent{25pt} \item \textbf{ALICE} will be able to store one year worth of data-taking on the disk buffer, which will, however, consist mostly of the few weeks of Pb--Pb data.
      The disk buffer contains all events in compressed raw data format, which is the same data stored permanently, and it writes with up to 100\,GB/s.}
\item \textbf{Processing}: Both experiments perform the calibration during the synchronous phase. LHCb reduces the data rate primarily via triggering while ALICE stores all events but performs data compression.
    {\setlength\itemindent{25pt} \item \textbf{LHCb} performs short term asynchronous processing soon after the data was stored to the buffer, and the buffer is overwritten after a short time.}
    {\setlength\itemindent{25pt} \item \textbf{ALICE} may run a short calibration phase following the Pb--Pb data-taking, and the remaining at least 10 months of the year are foreseen for asynchronous reconstruction.
      The compute budget will enable for up to two passes over the data if needed, and the disk buffer will be emptied thereafter for the next heavy-ion period after one year.
      Proton--proton data-taking will run in parallel or in between for few weeks.}
\end{compactitem}

\section{GPUs and FPGAs}

ALICE is the only experiment that uses GPUs in production, in the HLT during Run 2 and as the backbone of TPC reconstruction in Run 3.
All other experiments are studying GPU usage, in particular for their upgrade programs.
For Run 2, several prototypes have been developed, but not brought to production since the net improvements have been too small.
One big difference between ALICE and the other experiments is the TPC, which generates the majority of the data and causes the highest compute load while the reconstruction is parallelizable.
The TPC is thus ideally suited for GPU adoption.
In a more heterogeneous reconstruction scenario, more effort is required.
Common sense between all experiments is that the deployment of a significant number of GPUs is only worthwhile if they can be used steadily.
LHCb is investigating a possible GPU usage for the tracking in the HLT1.
Since the HLT1 will only run for less than~50\% of the time of a year, this raises concerns that the GPUs should not be idle in the HLT2.
ALICE needs GPUs to facilitate the synchronous reconstruction during Pb--Pb data-taking, which spans an even shorter period of time.
It is therefore essential to use as much of the GPU capacity as possible also during the asynchronous processing over the year.

All experiments are using FPGAs for the readout and in the hardware triggers.
On top of that, ALICE performs the TPC clustering on-the-fly with FPGAs saving many CPU cycles.
ATLAS has the most complex FPGA installation with the fast hardware tracker based on an associative memory.
In the same way as for GPUs, it is a waste of resources to have the compute power in FPGAs idle.
ATLAS plans to address this for the HTT tracker in Run 4 by using it for Monte Carlo productions when the LHC does not provide stable beam.
Overall, the GPU is still the more general option, which can be adopted with less effort for more reconstruction tasks.

\section{Conclusions}

After the upcoming upgrades for the LHC Run 3 and 4, all experiments ALICE, ATLAS, CMS, and LHCb will increase the event rate at least ten-fold.
The compute farms and storage capacities will grow, but this increase cannot compensate the increased data sizes.
Instead, online event reconstruction must improve and make better use of the available hardware, and the permanently stored data must be compressed stronger.
All experiments are at least evaluating the usage of hardware accelerators such as GPUs, while FPGA are used everywhere at least at the trigger level but also for certain parts of online reconstruction.
An important aspect is the usage of the accelerators in periods when there is no beam, which has implications on offline processing.
Instead of only storing raw data for later analyses, the trend goes towards more sophisticated online reconstruction.
ALICE and LHCb are building a system without hardware trigger processing the full data rate in software.
ATLAS and CMS face an even higher interaction rate and retain a hardware trigger but also extend the HLT farms.
The data rates in and out of the software farms are by large comparable, while ALICE features an asymmetric use case with the highest data rate but only during a short time of the year.

\end{document}